# The Prandtl Plus Scaling Approximation for Turbulent Boundary Layer Flows

By David W. Weyburne
*Air Force Research Laboratory, 2241 Avionics Circle, Wright-Patterson AFB, OH 45433, USA*

**Abstract**

Using the flow governing equation approach to similarity, Weyburne (D. Weyburne, arXiv:1701.02364, 2016) recently showed that for 2-D turbulent boundary layer flows, the Prandtl Plus scaling's are NOT, in general, the proper similarity scaling parameters. Based on that failure, Weyburne proposed alternative length and velocity scaling parameters that satisfy the flow governing approach to similarity for external turbulent boundary layer flows. Herein, we show that both the proposed new scaling's and the Prandtl Plus scaling's have a theoretical underpinning which requires that the boundary layer must show whole profile similarity for the scaling's to be applicable. The problem is that wall-bounded turbulent boundary layers are generally acknowledged as not showing whole profile similarity under any circumstances. Thus, applying scaling parameters that subsume whole profile similarity to boundary layers that do not show whole profile similarity means that, at best, the result represents a good approximation. Computer simulation results are used to explore these issues and provides at least preliminary experimental support of the theoretical failures.

## 1. INTRODUCTION

The so-called Logarithmic Law of the Wall is widely believed to describe the near wall behavior for all wall-bounded turbulent boundary layers. That is, there is a region of the turbulent boundary layer that is close (but not directly at the wall) that behaves logarithmically. According to Prandtl [1], in this region, the wall-bounded turbulent boundary layer velocity along the wall, $u(x,y)$, behaves as

$$\frac{u(x,y)}{u_\tau(x)} \cong \frac{1}{\kappa}\ln\left(\frac{y}{\nu/u_\tau(x)}\right)+C \quad , \tag{1}$$

where $u_\tau$ is the friction velocity, $\nu$ is the kinematic viscosity, $\kappa$ is the von Kármán constant, *C* is a constant, *x* is the flow direction along the wall, and *y* is the direction perpendicular to the wall. For reference purposes, the Prandtl Plus variables are $y^+ = yu_\tau/\nu$ and $u^+ = u(x,y)/u_\tau$ whereas the Prandtl Plus scaling's are $\nu/u_\tau$ and $u_\tau$. There has been an extensive research effort surrounding all aspects of the Logarithmic Law of the Wall. Recent reviews by Buschmann and Gad-el-Hak [2], Klewicki [3], and Maurusic, et al. [4] discuss the many current issues related to this topic.

Besides the basic scientific interest, the importance of the Log Law is that it is at the heart of many computational models for wall-bounded turbulent flows. As George [5] has pointed out, small changes in the $\kappa$ value in the turbulence models can have a large effect on the predicted drag. One would, therefore, expect that the Logarithmic Law of the Wall would have been experimentally verified to make up for the fact that there has never been a theoretical verification. However, the experimental verification will have to wait until new experimental techniques for measuring the friction velocity to higher accuracy are developed [4]. In spite of this lack of either experimental verification or theoretical verification, most in the flow community are convinced of the legitimacy of the Logarithmic Law of the Wall [2-7]. The only real controversy seems to be whether the von Kármán constant $\kappa$ is universal for external and internal flows [5,6].

Given the strong ongoing support for the Logarithmic Law of the Wall, it is not surprising that there has been no debate until recently about the proper scaling parameters for the near wall region. The Prandtl Plus scaling have universally been considered to be the appropriate length and velocity scales. Taking the velocity scale as the friction velocity is certainly plausible from a physics standpoint. The velocity gradient at the wall is the only experimentally accessible parameter for a typical boundary layer and is conveniently defined in terms of the so-called friction velocity as

$$\left.\frac{du(x,y)}{dy}\right|_{y=0} = \frac{u_\tau^2(x)}{\nu} \quad . \tag{2}$$

It is reasonable to expect that the velocity might scale with $\mu_\tau$ but what evidence do we have that the length scale in the near wall region behaves as $\nu / \mu_\tau$? The answer is that the only evidence would seem to be through the countless observations of Prandtl Plus scaled plots of turbulent boundary layer velocity profiles showing similar behavior in the Log Law region. However, this experimental observation based on "chi-by-eye" determination has never been backed up theoretically. In spite of a hundred years of effort, no one has ever proven that Prandtl Plus scaling's are the correct scaling's for the inner region of the wall-bounded turbulent boundary layer.

The theoretical landscape for the Prandtl Plus scaling has recently changed. Using the $\alpha$ and $\beta$ based Falkner-Skan momentum equation, Weyburne [8] showed that Prandtl Plus scaling's ONLY show similar behavior for the laminar and turbulent sink flow cases. The argument goes like this: if we use $\nu / \mu_\tau$ and $\mu_\tau$ as the scaling parameters in the $\alpha$ and $\beta$ based Falkner-Skan momentum equation, then one finds $\alpha = 0$ and that $\beta$ is given by a simple differential equation. The solution to the differential equation is that $\mu_\tau$ must behave as 1/(distance along the wall). This $\mu_\tau$ behavior together with $\alpha = 0$ is a characteristic of sink flow, flow in a converging/diverging channel. The Prandtl friction velocity based scalings, therefore, are NOT solutions to more general Falkner-Skan power law external flows.

By essentially reverse engineering the Prandtl Plus failure, Weyburne [8] introduced a new set of similarity scaling parameters for the inner region of the turbulent boundary layer velocity profile. The new parameters work for the both laminar and turbulent Falkner-Skan type flows. The new velocity scaling parameter, $u_0$, and the new length scaling parameter $\delta_0$, are defined as

$$\frac{u_0(x)}{\delta_0(x)} \propto \left.\frac{du(x,y)}{dy}\right|_{y=0} = \frac{u_\tau^2(x)}{\nu} \quad \Rightarrow \quad \delta_0 = \frac{a\nu u_0}{u_\tau^2} \quad , \tag{3}$$

where $a$ is a proportionality constant. This definition does not define the parameters to the point they can be evaluated separately. To do that Weyburne added a constraint in the form of the $\beta$ constraint of the $\alpha$ and $\beta$ based Falkner-Skan momentum equation. After substitution, this constraint becomes

$$a^2 u_0^2 \frac{du_0}{dx} = \frac{\beta u_\tau^4}{\nu} \quad . \tag{4}$$

where $\beta$ is a constant. It is still not possible to separate out $\delta_0$ or $u_0$ as stand-alone parameters. Hence, Eqs. 3 and 4 together are the formal definitions of the new scaling parameters. Preliminary experimental evidence [8] indicated that it is possible to collapse the inner region to a single curve using $\delta_0$ and $u_0$ in the same way that the Prandtl Plus scaling's have been shown to work.

In the work herein, we introduce a new similarity scaling rule which changes the theoretical landscape even further. In a paper dealing with inner region thickness, shape, and scaling parameters, Weyburne [9] recently pointed out that if similarity is present over the whole profile, it is possible to prove that the ratio of the length scale to the velocity scale MUST be proportional to the velocity gradient at the wall. If one looks at the Prandtl Plus scaling's and the new scaling's (Eq. 3), it is clear that that condition is met for both sets of parameters. The velocity gradient at the wall therefore formally expresses the relationship between the length and velocity scales. Now the problem for the wall-bounded turbulent boundary layer is that whole profile similarity has never been demonstrated using any scaling parameters [4,10]. Therefore, one cannot expect that both the length and velocity scales can be related by the velocity gradient at the wall. Due to the importance of this finding, we reproduce the relevant argument in the next Section.

The implications of this finding are far reaching. For one, it is apparent that the use of Eq. 1 is at best a good approximation. To test the new ideas, we used two Direct Numerical Simulation (DNS) data sets from Sillero, Jiménez, and Moser [11] and Komminaho and Skote [12]. The advantage of using the DNS data is that we know $u_\tau$ exactly so that the issues surrounding inter region scaling can be studied. Before we get to this testing work, we first reproduce the theoretical derivation for the length and velocity scaling relationship.

## 2. The Length and Velocity Scaling Relationship

### 2a. External Flows

In defining $u_0$ and $\delta_0$, Weyburne [8] imposed the restriction that the ratio of the length and velocity scaling parameters must be proportional to the square of the friction velocity. At the time, there was no solid theoretical justification for doing this. Recently, using an equal area integral similarity argument, Weyburne [9] showed that in fact if similarity is present in a set of profiles, then this is a necessary requirement. Both the Prandtl Plus scaling and the new scaling's satisfy this requirement. Due to the importance of this requirement to the work herein, we reproduce the derivation here since it makes it clear that this requirement is a whole profile similarity requirement.

The derivation starts with the definition of velocity profile similarity. According to Schlichting [13], a velocity profile at position $x_i$ is similar to the velocity profile at $x_j$ if

$$\frac{u(x_i, y/\delta_s(x_i))}{u_s(x_i)} = \frac{u(x_j, y/\delta_s(x_j))}{u_s(x_j)} \quad \text{for all y,} \tag{5}$$

where the length scaling parameter is $\delta_s(x)$, and the velocity scaling parameter is $u_s(x)$. Although the intent of Schlichting's Eq. 5 expression is clear, it appears to be based on redefining the velocity $u(x,y)$. In fact, the Schlichting's definition is a condensed way of expressing velocity profile similarity as

$$\frac{\overline{u}(x_i, y_i)}{u_s(x_i)} = \frac{\overline{u}(x_j, y_j)}{u_s(x_j)} \quad \text{where} \quad y_i = y_j, \quad \overline{u}(x_k, y_k) = u(x_k, y), \quad \text{and} \quad y_k = \frac{y}{\delta_s(x_k)} \quad \text{for all } y. \tag{6}$$

If similarity is present in a set of velocity profiles, then it is self-evident that the properly scaled second derivative profile curves (second derivative of Eq. 6) must also be similar. It is also self-evident that the area under the plotted scaled second derivative profile curves must be equal for similarity. In mathematical terms, the area under the scaled second derivative profile curve, $a(x)$, at $x_i$ is expressed by

$$\begin{aligned}
a(x_i) &= \int_0^{h_i} dy_i \; \frac{d^2 \left\{\overline{u}(x_i, y_i)/u_s(x_i)\right\}}{dy_i^2} \\
a(x_i) &= \frac{1}{u_s(x_i)} \int_0^{h_i} d\left\{\frac{y}{\delta_s(x_i)}\right\} \frac{d^2\overline{u}(x_i, y_i)}{d\left\{\dfrac{y}{\delta_s(x_i)}\right\}^2} = \frac{\delta_s(x_i)}{u_s(x_i)} \int_0^{h} dy \; \frac{d^2 u(x_i, y)}{dy^2} \\
a(x_i) &= \frac{\delta_s(x_i)}{u_s(x_i)} \left[\frac{du(x_i, y)}{dy}\right]_{y=h,0} \\
a(x_i) &= -\frac{\delta_s(x_i)}{u_s(x_i)} \left[\frac{du(x_i, y)}{dy}\right]_{y=0} \quad ,
\end{aligned} \tag{7}$$

where $h_i = h/\delta_s(x_i)$ and $h$ are located deep into the free stream above the wall. A necessary but not sufficient condition for similarity is that $a(x_1) = a(x_2)$. Using Eqs. 2 and 7, it is apparent that similarity requires

$$-\frac{\delta_s(x_1)}{u_s(x_1)} \frac{u_\tau^2(x_1)}{\nu} = -\frac{\delta_s(x_2)}{u_s(x_2)} \frac{u_\tau^2(x_2)}{\nu} = \text{constant} \quad . \tag{8}$$

The importance of this equation is that if similarity is present in a set of velocity profiles for any external 2-D wall bounded flow, then the ratio of the scaling parameters must be proportional to the square of the friction velocity.

## 2b. Internal Flows

The derivation above is applicable to external flows in that we assumed the upper bound was boundless. In this section, we look at the case of internal flows, in this case, channel flow. The arguments concerning similarity of the second derivative are identical. Starting with the definition for similarity (Eq. 6), it is self-evident that if similarity is present in a set of channel flow velocity profiles then the area under the scaled second derivative profiles must be equal. The area is given by

$$
\begin{aligned}
b(x_i) &= \int_0^{h_i} dy_i \, \frac{d^2\left\{\overline{u}(x_i,y_i)/u_s(x_i)\right\}}{dy_i^2} \\
b(x_i) &= \frac{1}{u_s(x_i)} \int_0^{h_i} d\left\{\frac{y}{\delta_s(x_i)}\right\} \frac{d^2\overline{u}(x_i,y_i)}{d\left\{\dfrac{y}{\delta_s(x_i)}\right\}^2} = \frac{\delta_s(x_i)}{u_s(x_i)} \int_0^{h} dy \, \frac{d^2u(x_i,y)}{dy^2} \\
b(x_i) &= \frac{\delta_s(x_i)}{u_s(x_i)} \left[\frac{du(x_i,y)}{dy}\right]_{y=h,0} \\
b(x_i) &= \frac{\delta_s(x_i)}{u_s(x_i)} \left( \left[\frac{du(x_i,y)}{dy}\right]_{y=h} - \left[\frac{du(x_i,y)}{dy}\right]_{y=0} \right) ,
\end{aligned}
\tag{9}
$$

where $h$ is now the channel height. A necessary condition for similarity is that the areas must be equal $b(x_1)=b(x_2)$ so that

$$
\begin{aligned}
\frac{\delta_s(x)}{u_s(x)} \left( \left[\frac{du(x_i,y)}{dy}\right]_{y=h} - \left[\frac{du(x_i,y)}{dy}\right]_{y=0} \right) &= \text{constant} \\
-2\frac{\delta_s(x)}{u_s(x)} \frac{u_\tau^2(x)}{\nu} &= \text{constant},
\end{aligned}
\tag{10}
$$

where we have flipped the sign of the top definition for the friction velocity to reflect the different orientation. This derivation assumes the top and bottom walls are identical. The importance of this equation is that if similarity is present in a set of velocity profiles for any internal 2-D wall bounded flow, then the ratio of the scaling parameters must be proportional to the square of the friction velocity.

## 2c. Implications

There are two important takeaways from the results presented in this Section. First, it applies to whole profile similarity. This is an important caveat for turbulent boundary layers since it is generally accepted that whole profile similarity is NOT possible for 2-D wall-bounded turbulent boundary layers [4,10]. The second important takeaway is that both the Prandtl Plus scaling and the new scaling satisfy this requirement. They satisfy the requirement but turbulent boundary layers do not show whole profile similarity. Therefore, this requirement should NOT be imposed on the length and velocity scaling parameters for wall-bounded turbulent boundary layer flows. Yet both the new scaling and the Prandtl Plus scaling impose this requirement. Therefore, while one can impose the requirement, one cannot expect the result to work correctly for flows that do not show whole profile similarity, i.e. wall-bounded turbulent boundary layer flows.

## 3. Determining the Friction Velocity using the New Scaling's

The theoretical results obtained thus far are perplexing. It appears that the neither the Prandtl Plus nor the new scaling's can be justified from a theoretical standpoint for wall-bounded turbulent boundary layer flows. In addition, the Prandtl Plus scaling should not work for general external wall-bounded flows. Yet, years of "chi-by-eye" observations indicate that the Prandtl Plus scaling's appears to work. What we conclude from this is that though unsupported theoretically, the Clauser Chart method for determining $u_\tau$ may be a close semi-empirical approximation under certain conditions. We would expect it to work well for near-equilibrium profiles, profiles that are close to being similar. If that is true, we reasoned the new parameters $u_0$ and $\delta_0$ may also work as a semi-empirical approximation.

To test this for the new scaling parameters we start by assuming Falkner-Skan [14] power-law type boundary layer behavior. If one assumes power-law behavior for the velocity scaling parameter and the friction velocity, it is easily verified that Eq. 3 and 4 are given by

$$u_0(x) = a_0 (x - x_0)^m \quad , \quad \delta_o(x) = \frac{a \nu a_0}{a_\tau^2} (x - x_0)^{(1-m)/2} \quad , \tag{9}$$

$$\text{and} \qquad u_\tau(x) = a_\tau (x - x_0)^{\frac{3m-1}{4}} \quad ,$$

which are the same as Falkner-Skan's original expressions for power-law laminar flows. Our proposed inner region scaling parameter test is based on measuring and analyzing the velocity profiles at a number of locations along the flow direction. To test the possibility that the friction velocity behaves as a power law flow, we turned to computer simulation data. The main advantage in using computer simulation data is that we know the friction velocity $u_\tau$ exactly. Hence, one can first verify that $u_\tau$ does follow power law behavior by fitting the $u_\tau$ using Eq. 9. Then we can test whether this same set of constants $m$ and $x_0$ found for the $u_\tau$ fit also work in $u_0$ and $\delta_0$ to produce similar behavior in the near wall region.

To do this we choose two direct numerical simulation (DNS) results from the literature that the authors have made available [11,12]. In each case, the direct numerically simulated ZPG data set consists of six or seven velocity profiles with supporting data including the friction velocity data $u_\tau$. In Figs. 1a and 1b we show the $u_\tau$ data fits to Eq. 9. For comparison purposes, we also show the Clauser Chart method [15] estimates of $u_\tau$. The Clauser Chart method used herein is implemented by least squares fitting of Eq. 1.

The next question is whether the fitted $m$ and $x_0$ values from Figs. 1a and 1b can result in the collapse of the velocity profile curves in the inner region. In Figs. 2a and 2b and Tables 1 and 2 we show the results for scaling the velocity profiles using the fitted $m$ and $x_0$ pairs. For comparison purposes, we show the $y^+$, $u^+$ profile plots in Fig. 3. It does appear that in the near wall region for the DNS fitted $m$ and $x_0$ pair there is a good collapse but that the collapse is very poor in the region traditionally associated with the Log Law. Even though the near wall collapse is better than the Clauser Chart method (Tables 1 and 2), the main point is that for the DNS fitted $m$ and $x_0$ pairs, the profile collapse does NOT extend into an experimentally accessible region.

## 4. Is there experimental confirmation for the Prandtl Plus Scaling problems?

The results shown in Figs. 1a and 1b indicate that while not perfect, the Clauser Chart method still appears to be able to estimate the $u_\tau$ values within 1-2% for ZPG near-equilibrium data sets. The DNS nature of these data sets lets us compare $u_\tau$ values obtained by the Clauser Chart method to exact values. Notice that the Clauser Chart method $u_\tau$ values in both cases are DC biased compared to the exact DNS values. We note that the bias is even worse (Tables 1 and 2) for the $\kappa$ =0.384, C=4.17 combination preferred by some. A second problem with the Clauser Chart method $u_\tau$ values for both data sets is that they have the wrong $du_\tau/dx$ behavior. A simple visual inspection confirms this claim. What could cause these problems? Some will argue that the Reynolds numbers are too low. That is certainly a possibility for Komminaho and Skote [12] data but should not be a problem for the Sillero, Jiménez, and Moser [11] data. If it is a Reynolds number problem, why do the two data sets DC bias behaviors look similar? While there is not a direct traceable connection to the afore mentioned theoretical problems, it is still significant that the Clauser Chart method does NOT reproduce the exact $u_\tau$ values for ZPG DNS data sets.

The next area we looked at was the idea that the length scale is not exactly $\nu/u_\tau$ as assumed in the Prandtl Plus scaling's. The advantage of the DNS data is that not only is there no traditional measurement noise, but the velocity profiles themselves are given to 9-15 significant digits. Thus, we are not limited to judging the profile collapse to using "chi-by-eye" observation of profile plots. Instead, we can look at the Normalized Sum of the Square of the Residuals (NSSR) where we use one of the profile plots themselves as the reference. To calculate the reported average NSSR value per point we start by averaging NSSR values from seven points per profile spanning the intended measurement range (either near wall, or Log Law region). The averaged NSSR values from each profile are then averaged together for the whole data set to get the final reported value. This gives a good measure of how well all the profiles in a set overlap one another. The low NSSR values in Tables 1 and 2 correspond to profile collapse that corresponds to about 3-4 significant digits overlap out of the 9-15 significant digits provided for *y* and *u*(*x*,*y*) by the authors. This indicates that there is a possibility for observing a considerable improvement in the overlap if we could find a better scaling set. To test this idea using the Prandtl Plus scaling, we tried a simple test. We assumed the velocity scale is $u_\tau$ and fixed it at the $u_\tau$ value obtained by the Clauser Chart method (least squares fitting to Eq. 1). The length scale was set initially starting with $\nu/u_\tau$ but these $u_\tau$ values are allowed to change to minimize the NSSR for the seven closest to the wall non-zero data points corresponding to $0 \le y^+ < 4$ for each profile. The actual averaged NSSR value per point for all six/seven profiles are given in Table 3. It is obvious that making relatively small changes in the length scale parameter can drastically improve the degree of profile overlap. The point here is that while the scaled $y^+$ and $u^+$ plots (Figs. 3a and 3b) look very good based on a "chi-by-eye" assessment, it is possible to do much better. However, such an improvement would not be easily detectable in a typical noise inhabited wind tunnel data set.

In doing the profile overlap check, the NSSR calculation involves linearly interpolating between scaled velocity values to ensure the comparisons are done on the same scaled *y*-value.

The Excel Forecast based one-line linear interpolation formula we used was very convenient but we had concerns that it might be introducing math errors into the least squares fitting operation. The first thing to note is that in the near wall region where we were working, $0 \le y^+ < 4$, the velocity profile is almost linear. Nevertheless, to check whether this was a problem, we did two separate tests where we individually fitted each of the first seven points of each profile to a seventh order polynomial with the intercept fixed at zero. The fits proved to be accurate to more than 14 digits. It then is a simple matter to recalculate the NSSR values using the polynomial fitted profiles. The results in both cases were within ~30% of the value calculated using the much simpler linear interpolation method. Given that the differences we observed (Table 3) are orders of magnitude, we are confident that the reported NSSR results are accurately reflecting the true situation.

## 5. Discussion

By far the most important issue uncovered in the above work is the fact that using a simple equal area integral method for similarity, we showed that $\delta_0$ and $u_0$ and the Prandtl Plus scaling's will NEVER correctly scale the typical wall-bounded turbulent boundary layer. They will only work for boundary layers displaying whole profile similarity. In the hundred years of research, not one turbulent boundary layer data set has ever been generated that displays whole profile similarity [4,10]. This scaling failure means that the Logarithmic Law of the Wall is at best a good approximation. This scaling failure means there is no chance that some von Kármán $\kappa$ value will be found that works, in general, for either internal or external wall-bounded turbulent boundary layers. The equal area integral method for similarity used to derive this result involves no approximations or assumptions. There is no other possible outcome unless one changes the definition of velocity profile similarity itself.

The problem comes down to the selection of the length scale appropriate to the near wall region of a turbulent boundary layer. Taking the velocity scale as the friction velocity is certainly plausible from a physics standpoint but to then take the length scale as $\nu / \mu_\tau$ is not so obvious. The only theoretical argument that this should be the case comes from the equal area similarity argument presented above. The problem, of course, is that it only applies to profiles showing whole profile similarity. This very argument means that there is not some yet to be discovered theoretical argument which will prove $\nu / \mu_\tau$ is the correct scaling for turbulent boundary layers not showing whole profile similarity. It is a case that based on one experimentally measured parameter (the velocity gradient at the wall), you can scale either the length or the velocity but not both. The successful application of one measured parameter to two scaling parameters requires some special circumstance for it to work, in this case, whole profile similarity. The experimental results presented herein support this argument. Table 3 NSSR values indicate that choosing a slightly different length scale can substantially improve the profiles overlap compared to the Prandtl Plus scaling. More DNS results, especially for flows will non-zero pressure gradients, will help to elucidate this matter further.

The other area of study above involves the new Falkner-Skan power law approximation for the near wall region. Although it appears that the method is not suited to finding $u_\tau$ values in experimental wind tunnel data sets, it certainly performs as well or better than the Clauser

Chart method in terms of scaling the near wall region (Tables 1 and 2), at least for ZPG DNS data sets. This makes it clear that there is nothing special about the Prandtl Plus scaling parameters. A different parameter set not discussed herein is described by Weyburne [9]. Although it also has the above problem, it has some advantages in describing the thickness and shape of the viscous inner region.

The implications of this scaling failure on the computational models for wall-bounded turbulent flows is not clear. These models are already purposely approximate in order to decrease the computational effort for complex geometries. The one area that will have to be reviewed is under what conditions one can expect the models to give accurate estimates. In light of the above results, it is certainly not the case that one can expect the models to work for every flow condition.

## 6. Conclusion

A new equal area similarity argument is used to prove that both the proposed new scaling's and the Prandtl Plus scaling's have a theoretical underpinning which requires that the boundary layer must show whole profile similarity for the scaling's to be applicable. Given that wall-bounded turbulent boundary layers are generally acknowledged as not showing whole profile similarity under any circumstances, this means the two parameter sets cannot be successfully applied to wall-bounded turbulent boundary layers. Computer simulation results were used explore these issues and provides at least some experimental support for the theoretical failures.

## Acknowledgement

The author acknowledges the support of the Air Force Research Laboratory and Gernot Pomrenke at AFOSR. In addition, the author thanks the experimentalists for making their data sets available for inclusion in this manuscript.

## Tables and Figures

**Table 1: Sillero, Jiménez, & Moser [11] Fitted Values, Near Wall**

| $m$ | $x_0$ | NSSR/NSSR$_0$ of Near Wall* | Profiles Collapsed? | Average Deviation of $u_\tau$** |
|---|---|---|---|---|
| 0.2393 | -153.8 | 1 | Only near wall | 0.044% |
| $y^+$, $u^+$ | $\kappa$=0.41, C=5 | 1.06 | Yes | 1.0% |
| $y^+$, $u^+$ | $\kappa$=0.384, C=4.17 | 1.6 | Yes | 1.6% |

*The normalized sum of the squares of the residuals (NSSR) per data point is calculated using the first seven non-zero data points from each curve. NSSR$_0$ is the average NSSR value per data point for the $m$=0.2393 fit for the seven data points nearest the wall. The NSSR is calculated relative to the scaled reference profile arbitrarily chosen to be $Re_\theta$=5000.

**The average deviation is calculated using the shown $m$ and $x_0$ values in Eq. 9 and then least squares fitting for $a_\tau$ using the exact $u_\tau$ values as the reference. With $a_\tau$, $m$, and $x_0$ values, the $u_\tau$ fitted value is calculated and compared to the exact value as the average deviation. The value for $y+$, $u+$ is the Clauser Chart value compared to the DNS actual value.

**Table 2: Komminaho and Skote [12] Fitted Values, Near Wall**

| $m$ | $x_0$ | NSSR/NSSR$_0$ of Inner Region* | Profiles Collapsed? | Average Deviation of $u_\tau$** |
|---|---|---|---|---|
| 0.24895 | 65.05 | 1 | Only near wall | 0.1% |
| $y+$, $u+$ | $\kappa$=0.41, C=5 | 8 | Yes | 1.3% |
| $y^+$, $u^+$ | $\kappa$=0.384, C=4.17 | 6 | Yes | 2.7% |

* The normalized sum of the squares of the residuals (NSSR) per data point is calculated using the first seven non-zero data points from each curve. NSSR$_0$ is the average NSSR value per data point for the $m$=0.24895 scaling's seven data points closest to the wall. The NSSR is calculated relative to the scaled reference profile ($Re_\theta$=499).

**The average deviation is calculated using the shown $m$ and $x_0$ values in Eq. 9 and then least squares fitting for $a_\tau$ using the exact $u_\tau$ values as the reference. With $a_\tau$, $m$, and $x_0$ values, the $u_\tau$ fitted value is calculated and compared to the exact value as the average deviation. The value for $y+$, $u+$ is the Clauser Chart value compared to the DNS actual value.

**Table 3: Profile Collapse Test for Different Length Scales**

| | $u_\tau$* condition | NSSR of Near Wall | NSSR Ratio | Average Difference of $u_\tau$** |
|---|---|---|---|---|
| Sillero, Jiménez, & Moser [11] | $u_\tau$ same | 4.7E-6 | 470 | - |
| Sillero, Jiménez, & Moser [11] | $u_\tau$ different | 1.0E-8 | 1 | 0.15% |
| Komminaho & Skote [12] | $u_\tau$ same | 8.6E-5 | 10,230 | - |
| Komminaho & Skote [12] | $u_\tau$ different | 8.4E-9 | 1 | 1.0% |

* The $u_\tau$ condition "$u_\tau$ same" means the Clauser Chart method $u_\tau$ values are used for both the length scale and the velocity scale. The $u_\tau$ condition "$u_\tau$ different" means the Clauser Chart method $u_\tau$ values are used for the velocity scale but the length scale is the least squares fitted value obtained using NSSR per data point calculated for the first seven non-zero data points from each curve.

**The average difference is calculated using the least squares fitted length scale $u_\tau$ values compared to the Clauser Chart method $u_\tau$ values.

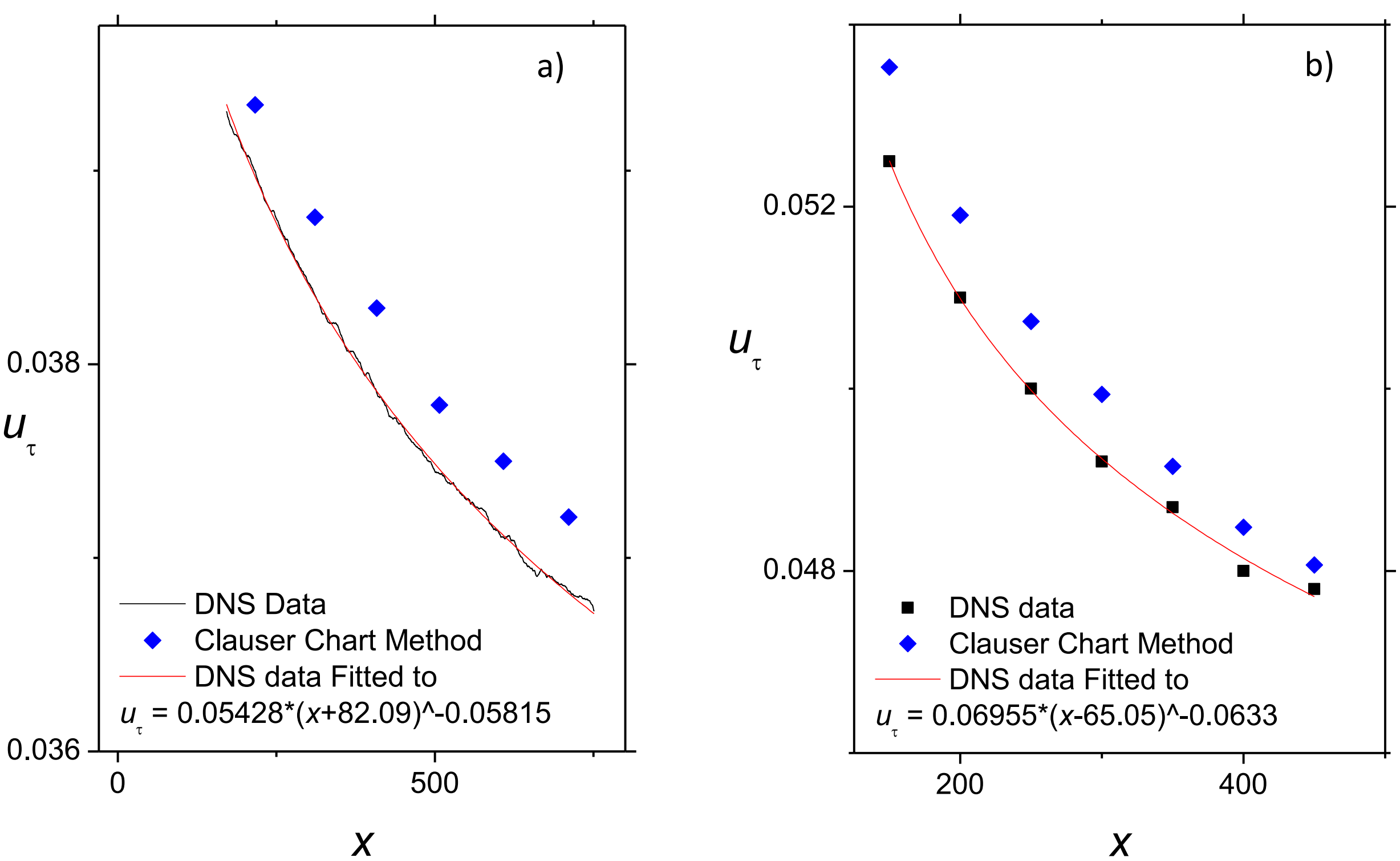

Fig. 1: The DNS $u_\tau$ values are in **black** and the red lines are the fitted lines (Eq. 9). Figure a) the Sillero, Jiménez, and Moser's [11] data set and b) the Komminaho and Skote's [12] data set. The ♦ are calculated by the Clauser Chart method, κ=0.41 and C=5.

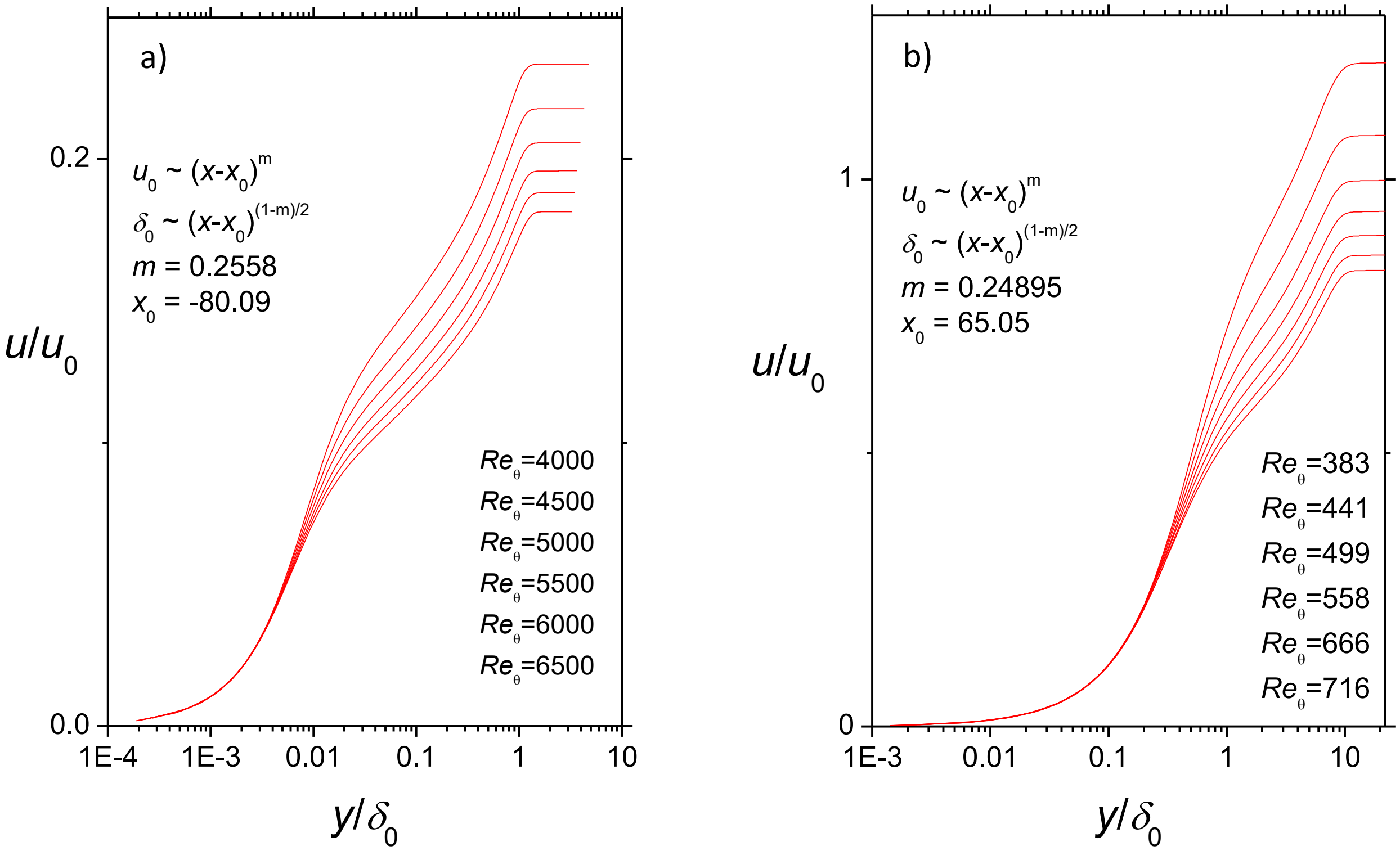

Fig. 2: The scaled profile data plotted using $\delta_0$ and $u_0$. The a) Sillero, Jiménez, and Moser's [11] data set, and b) the Komminaho and Skote's [12] data set. Collapse occurs in the near wall region.

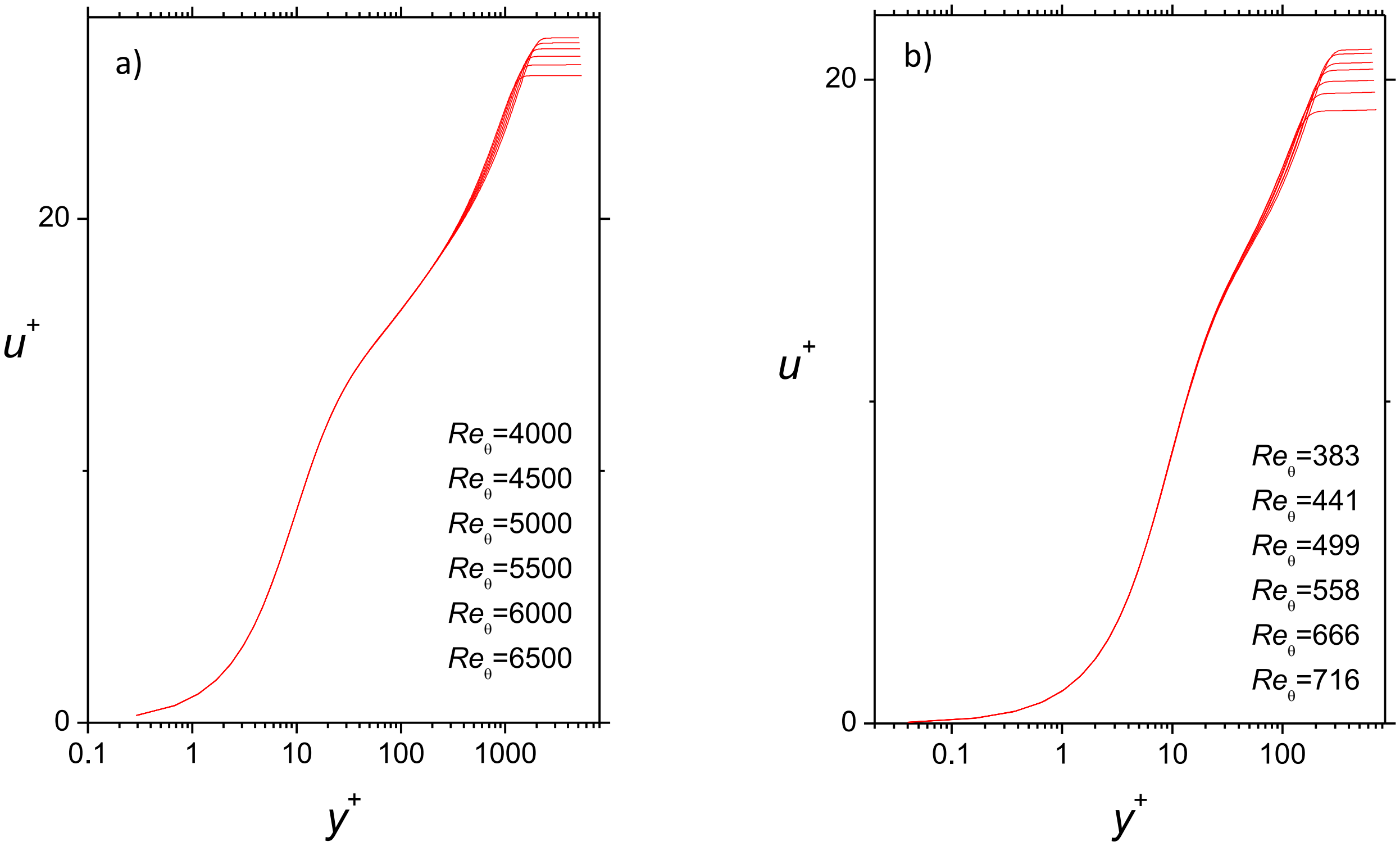

Fig. 3: The Prandtl Plus scaled profile data. a) The Sillero, Jiménez, and Moser's [11] data set, and b) the Komminaho and Skote's [12] data set. Collapse occurs into the Log Law region.